# Recovering the flat-plane condition in electronic structure theory at semi-local DFT cost

Akash Bajaj[1,2], Jon Paul Janet[1], and Heather J. Kulik[1,*]

[1]*Department of Chemical Engineering, Massachusetts Institute of Technology, Cambridge, MA 02139*

[2]*Department of Materials Science and Engineering, Massachusetts Institute of Technology, Cambridge, MA 02139*

AUTHOR INFORMATION

**Corresponding Author**

*email: hjkulik@mit.edu, phone: 617-253-4584

**ABSTRACT**: The flat plane condition is the union of two exact constraints in electronic structure theory: i) energetic piecewise linearity with fractional electron removal or addition and ii) invariant energetics with change in electron spin in a half filled orbital. Semi-local density functional theory (DFT) fails to recover the flat plane, exhibiting convex fractional charge errors (FCE) and concave fractional spin errors (FSE) that are related to delocalization and static correlation errors. We previously showed that DFT+U eliminates FCE but now demonstrate that, like other widely employed corrections (i.e., Hartree Fock exchange), it worsens FSE. To find an alternative strategy, we examine the shape of semi-local DFT deviations from the exact flat plane, and we find this shape to be remarkably consistent across ions and molecules. We introduce the jmDFT approach, wherein corrections are constructed from few-parameter, low-order, functional forms that fit the shape of semi-local DFT errors. We select one such physically intuitive form and incorporate it self-consistently to correct semi-local DFT. We demonstrate on model systems that this jmDFT approach represents the first easy-to-implement, no-overhead approach to recovering the flat plane from semi-local DFT.

Approximate density functional theory (DFT) is one of the most widely-used electronic structure methods, despite the fact that presently available pure exchange-correlation (xc) approximations in Kohn-Sham DFT are plagued by both one- and many-electron self-interaction errors (SIE)[1-5], which give rise to well-known problems in dissociation energies[2, 6-9], barrier heights[10], band gaps[11-12], electron affinities[13-15], and other manifestations of delocalization error[16-18]. A number of generalizations to Kohn-Sham DFT, e.g., range-separated[19-25] and global hybrids as well as DFT+U[26-29], derive their error mitigating properties from recovering the piecewise linearity of the energy with respect to fractional electron removal or addition, a requirement of any exact electronic structure method[30-32]. The combination of the requirement of piecewise linearity with the additional constraint that the energy is invariant with respect to the spin of an electron in isoenergetic orbitals produces what has come to be known as the flat-plane condition[33-34] in which two planes meet at a fractional spin line[35] (FSL) seam.

The flat plane condition applies to any three electronic states with $N$-1, $N$, and $N$+1 electrons, where i) an orbital is half-occupied, doubly-occupied, and empty in the $N$-, $N$+1-, and $N$-1-electron states, respectively, and ii) the $N$-electron multiplicity is thus higher by one than for the other two states. Ionization potentials (IPs) or electron affinities (EAs) connecting these electronic states, which are available for ions from experimental databases[36], enable construction of "exact" experimental flat planes. An exhaustive study[37] of such flat planes has revealed that nearly all feasible electronic state combinations preserve this shape, with a few exceptions for highly ionized and excited electronic states outside the scope of the current work. Our focus on ground state DFT narrows our study to cases where all three electronic states correspond to the ground state of the ion or to the lowest state with that quantum number: ionization of i) *s* electrons in H-Rb or He$^+$-Sr$^+$ ($N$-1/$N$+1: $^1$S, $N$: $^2$S), ii) *p* electrons in N-Bi or O$^+$-Te$^+$ ($N$-1/$N$+1: $^3$P, $N$: ⁴S), iii) $d$ electrons in $Fe^{3+}$ or $Co^{4+}$ ($N$-$1$/$N$+$1$: ⁵D, $N$: ⁶S) (Supporting Information Text S1 and Table S1). A similar approach is used to construct molecular ($H_2$, $Na_2$) flat planes, albeit with correlated wavefunction theory reference IP and EA data (Supporting Information Text S1-S2).

We recently demonstrated[38] that DFT+U[26-29] recovers piecewise linearity[30-32] and the derivative discontinuity[30] in representative transition metal ions and inorganic complexes. The most widely applied DFT+U correction[26-29, 39] is:

$$E^{DFT+U} = E^{DFT} + \frac{1}{2}\sum_{I,\sigma}\sum_{nl} U_{nl}^I \left[ \text{Tr}\left(\mathbf{n}_{nl}^{I,\sigma}\left(\mathbf{1}-\mathbf{n}_{nl}^{I,\sigma}\right)\right)\right] \quad (1)$$

where $U_{nl}^I$ is applied to the $nl$ subshell of the $I$th atom, $\sigma$ is the spin index, and $\mathbf{n}_{nl}^{I,\sigma}$ is the occupation matrix with elements, $n_{mm'}^{I,\sigma} = \sum_{k,v}\langle\psi_{k,v}|\phi_{m'}^I\rangle\langle\phi_m^I|\psi_{k,v}\rangle$, obtained from projecting extended states, $\psi$, onto atomic orbitals, $\phi$. DFT+U thus acts as a local correction on valence orbitals with strong $nl$ character. In the eq. 1 approximation, differences in Coulomb repulsion between same- and opposite-spin electrons (i.e., exchange) have been neglected, and, $U_{nl}^I$ is an effective $U_{eff}$=$U$-$J$[28-29, 39]. Even in the atomic limit where $\delta N \sim \delta n$, we noted that piecewise linearity is not recovered at linear-response $U_{eff}$[39-40], i.e. $\frac{\partial^2 E}{\partial n^2}$, but instead when the value corresponding to average curvature obtained from frontier orbitals[41], i.e., $\left\langle\frac{\partial^2 E}{\partial N^2}\right\rangle$, is applied.

By construction, this DFT+U formulation should dramatically increase energetic errors along the FSL (i.e., $n_\alpha$+$n_\beta$=1, where $n_\alpha$ and $n_\beta$ are the fractional electron counts of α and β orbitals emptying or filling in the flat plane). As an example, recovering piecewise linearity with DFT+U with fractional electron addition from He$^+$ to He (i.e., the fractional charge line, FCL) requires a $U_{\text{eff}} = \left\langle \frac{\partial^2 E}{\partial N^2} \right\rangle$ = 16 eV (Figure 1). Examining errors along the He$^+$ FSL reveals that modest errors for PBE (max. value ca. 2 eV) are dramatically (i.e., 200%) increased to 6 eV with DFT+U (Figure 1). Similar increases in FSL errors are known to occur for methods that incorporate Hartree-Fock (HF) exchange, e.g., full HF, global hybrids, or range-separated hybrids[35, 42]. Such a large increase in fractional spin error has been largely overlooked in the application of tuned range-separated hybrids for organic molecules[43-48] and some inorganic[49-50] systems, with few exceptions[42]. Nevertheless, it has been known for some time that increasing FSL error is directly related to increasing static correlation errors (SCE)[35]. Thus, this analysis highlights that, like incorporation of HF exchange, DFT+U directly increases SCE, which should be particularly concerning for already-SCE-prone transition metal complexes and solids to which DFT+U is widely applied.

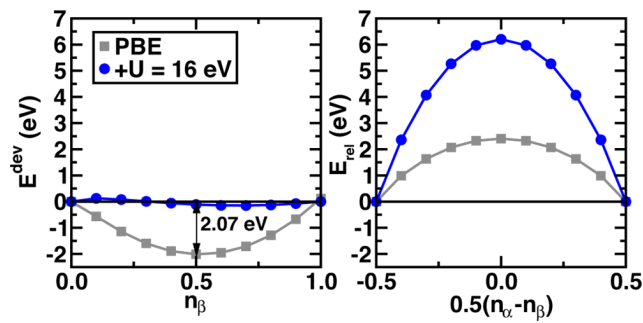

**Figure 1.** (Left) Deviation ($E^{\text{dev}}$, in eV) from linearity of PBE (gray squares) and after inclusion of $U$ = 16 eV (blue circles) for the He$^+$ to He fractional charge line. The maximum deviation from linearity of PBE is annotated. (Right) Relative energy ($E_{\text{rel}}$, in eV) along the fractional spin line of He$^+$ for PBE and with $U$ = 16 eV. Lines correspond to interpolation of the explicitly calculated points.

Thus, we depart from traditionally applied methods to identify if other functional forms can recover piecewise linearity and the derivative discontinuity while decreasing or even eliminating FSL errors. We consider functional forms that are easy to implement as self-consistent corrections to the total energy and the potential within the context of a Kohn-Sham DFT calculation without increasing computational cost. In order to identify such functional forms, we return to the flat-plane condition and evaluate the deviations of semi-local DFT functionals (here, PBE[51], similar behavior from other xcs shown in Supporting Information Figure S1) from the exact flat-plane condition on ions and molecules. We choose He$^+$ as our primary worked example as observations are transferable to other elements but the large PBE errors make it a challenging test case (Figure 2 and Supporting Information Text S1-S4 and Tables S1-S14).

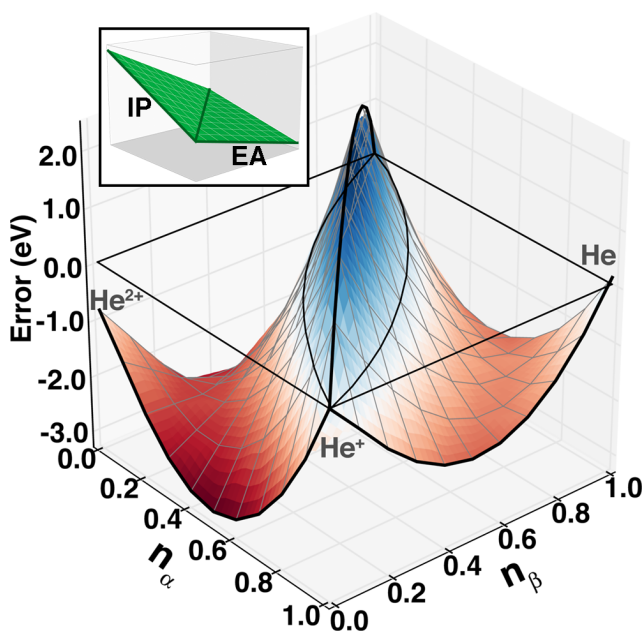

**Figure 2.** Deviation (i.e., error) of the PBE functional (in eV) from the flat plane condition constructed from the experimental IP and EA (schematic shown in inset). The plane is colored according to the sign of errors (from negative in red through white to positive in blue). Endpoint charge states are annotated in gray.

The errors in PBE with respect to the exact flat plane are convex along the FCL and concave along the FSL (Figure 2 and Supporting Information Figure S2). Examination of this error also i) suggests that the smoothly varying deviations are likely to be suitably fit with a low-order polynomial of $n_\alpha$ and $n_\beta$ in the spirit of DFT+U and ii) highlights why a penalty on fractional occupations (eq. 1) can correct the edges of the flat plane (i.e., the FCLs) but will dome the middle region, exacerbating FSL errors (Figure 2). Advantages of low-order polynomial corrections are that i) they add no overhead in the solid state, and ii) they are transparent, enabling the full derivation of their behavior over a range of expected electron configurations[38], as opposed to more opaque many-parameter approximate functionals that can unpredictably lack good extrapolative performance on molecules that differ from training data.

First focusing on the FSL error, it is clear that a second-order expression is needed that i) opposes the intrinsic PBE penalty for $n_\alpha=1/2$, $n_\beta=1/2$ with respect to an $n_\alpha=1$ or $n_\beta=1$ configuration and ii) approaches zero as $n_\alpha$ or $n_\beta$ approach 1 (Figures 1-2). A functional form that can eliminate this FSL error is the $J$ correction in DFT+U+J:

$$E^{\text{DFT+U}} = E^{\text{DFT}} + \frac{1}{2}\sum_{I,\sigma}\sum_{nl}(U_{nl}^I - J_{nl}^I)\left[\text{Tr}\left(\mathbf{n}_{nl}^{I,\sigma}\left(\mathbf{1} - \mathbf{n}_{nl}^{I,\sigma}\right)\right)\right] + \frac{1}{2}\sum_{I,\sigma}\sum_{nl}J_{nl}^I\left[\text{Tr}\left(\mathbf{n}_{nl}^{I,\sigma}\mathbf{n}_{nl}^{I,-\sigma}\right)\right] \quad (2)$$

If we apply eq. 2 to only correct the FSL in He$^+$, we would thus require a negative $J$ of -16 eV and $U_{\text{eff}}=U-J=0$. There is a key reason why DFT+U+J alone cannot recover the flat plane condition that should be relevant to researchers wishing to broadly employ the method for correlated molecules: the $J$ term in eq. 2 increases with increasing filling of the $nl$ subshell, producing an arbitrary shift in the energy of differing oxidation states (Supporting Information Figure S3). The primary assumption in most functional tuning approaches, consistent with our own data, is that integer-electron endpoint errors are modest and endpoint adjusting terms (e.g., +*J*) are not beneficial in most cases where the magnitude of the needed correction is not known *a priori* (Supporting Information Table S1).

Thus, we now consider generalizations of DFT+U+J that can be used to construct self-consistent corrections to semi-local functionals on the basis of the shape of the error, much as a jello mold is used to create a 3D shape. The resulting form can then be incorporated self-consistently to counteract such errors and recover the flat plane. Working with ions and molecules (e.g., $H_2$, $Na_2$), we use as the input to our fit the difference between semi-local DFT and exact flat plane energies along a grid of $n_\alpha$ and $n_\beta$ values for the emptying or filling electron level. We judge these corrections by how well they fit to the shape of the error, through root mean square (RMSE) measures and by the number of new or established (e.g., *U*, *J*) parameters they introduce and assume changes in electron count will be reflected in occupations. Our approach, which we refer to in its broadest form as jello mold DFT (jmDFT), no longer requires a grounding in the Hubbard model in its original form[52] or extensions[53-56] to construct model potentials. Consistent with earlier observations, DFT+U or DFT+U+J terms make poor fits to the PBE errors: the best fits are to DFT+U+J with differing U and J values (i.e., 4 parameters) at around 0.1-0.5 eV RMSE, and DFT+U-only errors are much larger (Supporting Information Tables S3-S14).

Examining shapes and parameters of fit expressions across elements confirms that errors are generally the same shape but this also leads to observations consistent with earlier work on only delocalization error[38, 44, 57-60]: heavier elements exhibit reduced convexity and concavity as the addition or removal of a single electron becomes a smaller fraction of total electron count (i.e., smaller coefficients in fits, Supporting Information Tables S3-S14).

The most general fit can obtain an order of magnitude lower RMSE (ca. 0.01-0.05 eV overall, 0.08 eV in He, Figure 3) than DFT+U-based approaches by applying a second-order polynomial in the electron count of orbitals that empty and fill between states in the flat plane, $n_\alpha$ and $n_\beta$:

$$\Delta E = a + b\left(v_1 - \frac{1}{2}\right)^2 + c(v_2) + d(v_2)^2 = a + \frac{b}{4}(n_\alpha - n_\beta)^2 + \frac{c}{2}(n_\alpha + n_\beta - 1) + \frac{d}{4}(n_\alpha + n_\beta - 1)^2 \quad (3)$$

This functional form arises from observing that transforming the flat plane error to a coordinate system $v_1=1/2(n_\alpha-n_\beta+1)$ and $v_2=1/2(n_\alpha+n_\beta-1)$ aligns the FSL to $v_2=0$ and reveals the directions of approximately parabolic dependence (Supporting Information Text S3). Rather than fitting the three extrema with a fourth order polynomial with up to sixteen coefficients, we employ asymmetric coefficients around the FSL (i.e., 8 parameters) in eq. 3 to recover the FSL ridge (Supporting Information Text S3). Although this leads to a good fit, the large number and non-intuitive nature of these parameters, including linear terms[61] and constants, leads us to disfavor this correction form. This most general form bears no connection to DFT+U+J but grouping terms reveals $U$- or $J$-like components (Supporting Information Text S3).

We return to DFT+U+J to identify additions and adjustments that can recover the flat plane. We first add a third term, $K$, that counteracts energy increases from $J$ with orbital filling. The $K$ fitting coefficients will be zero for the $N$-1- to $N$-electron portion of the flat plane but counteract the $J$ term on the $N$- to $N$+1-electron side ($He^{2+}$ to $He^+$: $K$=0.6 eV, $He^+$ to He: $K$=-25.5 eV):

$$E^K = \frac{1}{2}\sum_{I,\sigma}\sum_{nl} K_{nl}^I \left[ \text{Tr}\left((\mathbf{n}_{nl}^{I,\sigma} - \mathbf{n}_{nl}^{I,-\sigma} + 1)(\mathbf{n}_{nl}^{I,-\sigma} - \mathbf{n}_{nl}^{I,\sigma} + 1)\right)\right] \quad (4)$$

Physically, eq. 4 could provide a finer tuning of spin state ordering than DFT+U alone by having no effect on intermediate spin states but favoring low or high-spin states, if $K$ is negative or positive, respectively. Asymmetric DFT+U+J+K introduces six parameters and relies on the difference of two large effects (i.e., +J and +K) to cancel, introducing precision or stability issues but otherwise giving a more initiutive fit with comparable quality to the polynomial (Figure 3, asym. RMSE in He: 0.11 eV, 0.01-0.05 eV for heavier elements, see Supporting Information Tables S3-S14 and Figure S4). Alternatively, we also adjust the J term in eq. 2 ($U+J/J'$), resetting the electron count on the $N$- to $N+1$-electron side:

$$E^{J'} = \frac{1}{2}\sum_{I,\sigma}\sum_{nl} J'^I_{nl}\left[\text{Tr}\left((\mathbf{n}_{nl}^{I,\sigma} - 1)(\mathbf{n}_{nl}^{I,-\sigma} - 1)\right)\right] \quad (5)$$

retaining eq. 2 for the $N$-1- to $N$-electron side. Although the difference in curvature on the FCL is known to typically require different parameters for different ionization processes[38, 62], the U+J/J' form appears to permit reduction in the number of parameters: symmetric coefficients (i.e., $J=J'$, 2 parameters) are of comparable quality to asymmetric (4 parameters, see Supporting Information Tables S3-S14). The symmetric coefficient model provides among the lowest RMSEs (ca. 0.03 eV in most $s$-block elements and 0.10 eV in most $p$-block elements, higher in He: sym. RMSE = 0.35 eV, asym. 0.15 eV) (Figure 3 and see Supporting Information Tables S3-S14). Here, $U$ is positive, whereas $J$ is negative, and increasing atomic number decreases coefficients (i.e., He sym. U+J/J' $U = 20.4$ and $J= -30.4$ eV is $U = 4.8$ and $J = -5.7$ eV for Ca, see Supporting Information Table S5). Alternative functional forms considered were not pursued further because they introduced more parameters or did not improve upon standard DFT+U+J (Supporting Information Text S3 and Tables S3-S14).

Decomposing total errors into contributions from different regions of the flat plane quantitatively confirms earlier observations on four representative fits for He (Figure 3). The lowest errors observed in He$^+$ with the polynomial are achieved through balanced error elimination in all regions of the plane but especially eliminating FSL errors (RMSE = 0.08 eV). Conversely, $U+J$ produces large errors there and in the He$^+$ to He FCL (RMSE = 0.56 eV) that are eliminated with $U+J/J'$ (RMSE = 0.15 eV, Figure 3). For the He example, the $U+J/J'$ RMSE (0.15 eV) remains larger than the 6-parameter $U+J+K$ (RMSE = 0.11 eV) and polynomial (RMSE = 0.08 eV) correction because these approaches correct integer electron errors, improving FCL+ (PBE He$^+$ IP error of 0.85 eV, Figure 3a). The three forms are more comparable in other elements that have smaller endpoint error (Supporting Information Tables S3-S14).

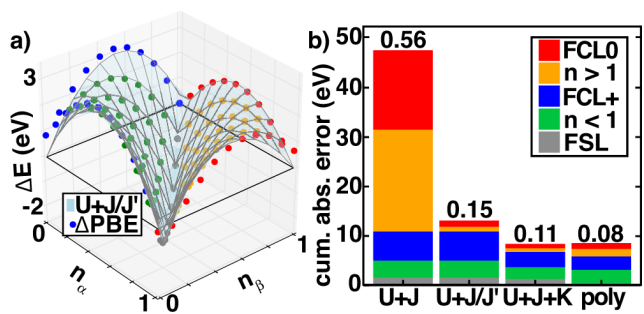

**Figure 3.** a) Needed correction (ΔPBE, in eV) to recover the flat plane condition from PBE indicated as dots colored according to their definition in legend shown in b along with the U+J/J' correction shown as a translucent blue surface. b) The cumulative absolute error over the entire plane for four fitting schemes, U+J, U+J/J' (shown in a), U+J+K, and polynomial (poly). The stacked bar is annotated by the RMSE error of the fit in eV, and points are grouped into i) fractional charge line of He$^+$ to He (FCL0, red), ii) cases on the $n_\alpha+n_\beta > 1$ side of the plane not otherwise specified (orange), iii) fractional charge line of He$^{2+}$ to He$^+$ (FCL+, blue), iv) cases on the $n_\alpha+n_\beta < 1$ side of the plane not otherwise specified (green), and the fractional spin line (gray). Each point is indicate for the U+J/J' case at left.

At this point we select the U+J/J' correction as the functional form we will employ in our jmDFT approach to build a fully self-consistent functional that can be applied across the periodic table. We chose this form because in several cases (e.g., *s*-block Be, Mg, Ca, or *p*-block S, Se, P, As), the symmetric, 2 parameter fit is of comparable fit to the asymmetric, 4 parameter one and it does not introduce an adjustable parameter to the total energy at integer electron number (Supporting Information Tables S3-S14). The energy expression in eqs. 2 and 5 is incorporated self-consistently by inclusion of a potential correction of the form:

$$V^{\mathrm{jmDFT}} = \sum_{I,\sigma} \sum_{nl} \sum_{m} \left( \frac{U_{nl}^{I}}{2}\left(1 - 2n_{nl,m}^{I,\sigma}\right) + J\left(n_{nl,m}^{I,-\sigma} - 1\right) \right) \left|\phi_{nl,m}^{I}\right\rangle\left\langle\phi_{nl,m}^{I}\right| \qquad (6)$$

where the 1 in the second parentheses is only present for the J' case. We have added both this correction and the *U+J+K* form into locally-modified copies of Quantum-ESPRESSO[63], source code available upon request. Incorporating this new potential and energy correction introduces no overhead or complexity beyond the ingredients of widely employed semi-local DFT corrections, i.e., we require that occupations obtained from projecting onto atomic states remain a good proxy for the fractional electron count that we used in the initial fit (i.e., $\delta N \sim \delta n$). We recently demonstrated the range of applicability of this assumption in larger complexes, and alternate forms, currently being developed by our group, could be employed more suitably in the rare cases where this approximation fails[38]. Self-consistent evaluation of the energy with the jmDFT potential correction satisfied this requirement at all fractional charge and spin points for our He example (other examples in Supporting Information Figures S5-S9). The final plot of the energy versus $n_\alpha$ and $n_\beta$ electron filling reveals a recovered flat plane (Figure 4). This flat plane includes the seam at the FSL and a derivative discontinuity at the meeting point between the two

FCLs (Figure 4). Examining the FSL and FCL from $He^+$ to He reveals that errors obtained with PBE are dramatically reduced with respect to the exact answer for He. Evaluation on a number of other *s* electron systems reveals the same result, with the only difference being the magnitude of the flat plane deviations to correct, which generally decreases with increasing electron number (Supporting Information Figures S5-S8).

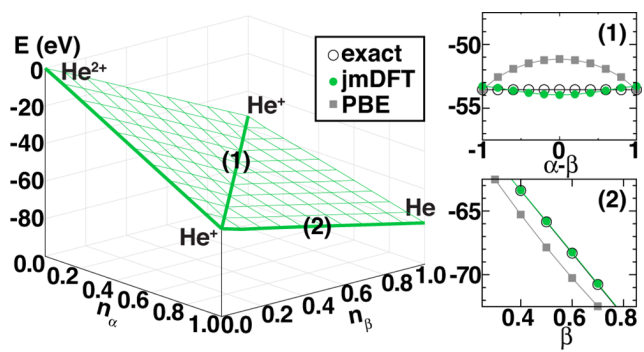

**Figure 4.** (Left) Total energies with jmDFT that recover the flat plane annotated with two points (1, FSL and 2, FCL0) for insets at right. (Right, top) The FSL for jmDFT (green circles), PBE (gray squares), and the exact result (open circles), with legend shown at left. (Right, bottom) The FCL0 for jmDFT, PBE, and the exact result, with legend same as top right and legend shown at left.

In conclusion, we have introduced a practical, no-overhead solution to recovering the flat plane within semi-local DFT. This general jmDFT framework involves fitting to known errors as a function of electron occupation or spin and incorporating the resulting form as a self-consistent correction that adds no computational cost. We have identified that appropriate functional forms are remarkably transferable from *s* to *p* to *d* electron atoms and ions as well as to small molecules, as judged through similarly low RMSEs for expressions that capture FCL convexity and FSL concavity. This result suggests that one general functional form should be suitable to recover exact conditions in a wide range of small molecules. We have selected a form reminiscent of Hubbard model corrections because it offers the best balance of the fewest number of parameters with physical transparency in the nature of the correction and as we extend our methodology to larger molecules and solids, we anticipate that the best choices for the adjustable components in this methodology, such as the form of the projections and the chosen polynomial correction, could be different. Finally, we expect that these functional forms may help guide alternate approaches to more directly tackle static correlation error minimization and piecewise linear recovery in DFT.

**ACKNOWLEDGMENT**

The authors acknowledge support by the Department of Energy under grant number DE-SC0018096, the Office of Naval Research under grant number N00014-17-1-2956, and MIT Energy Initiative seed grants (2014, 2017). J.P.J. was supported in part by an MIT-SUTD Graduate fellowship. H.J.K. holds a Career Award at the Scientific Interface from the Burroughs Wellcome Fund. This work was carried out in part using computational resources from the Extreme Science and Engineering Discovery Environment (XSEDE), which is supported by National Science Foundation grant number ACI-1053575. The authors thank Adam H. Steeves for providing a critical reading of the manuscript.